\documentclass[
reprint,
groupedaddress,
 amsmath,amssymb,
 aps,
floatfix,
]{revtex4-2}

\usepackage{graphicx}% Include figure files
\usepackage{dcolumn}% Align table columns on decimal point
\usepackage{bm}% bold math
\usepackage[caption=false]{subfig}
\usepackage[colorlinks,citecolor=blue]{hyperref}
\usepackage{float}
\usepackage[mathlines]{lineno}% Enable numbering of text and display math
\begin{document}

\preprint{APS/123-QED}

\title[{Kinetic characteristics of ions in an inertial electrostatic confinement device}]{Kinetic characteristics of ions in an inertial electrostatic confinement device}% Force line breaks with \\
%\thanks{}%

\author{D. Bhattacharjee}
\affiliation{Center of Plasma Physics - Institute for Plasma Research, Sonapur, Kamrup-782402, India}
\author{S. Adhikari}
\affiliation{Department of Physics, University of Oslo, PO Box 1048 Blindern, NO-0316 Oslo, Norway}

\author{N. Buzarbaruah}
\affiliation{Center of Plasma Physics - Institute for Plasma Research, Sonapur, Kamrup-782402, India}

\author{S. R. Mohanty}%
\altaffiliation[Also at ]{Homi Bhabha National Institute, Anushaktinagar, Mumbai, Maharashtra, 400094, India}
\affiliation{Center of Plasma Physics - Institute for Plasma Research, Sonapur, Kamrup-782402, India}
\email{smruti@cppipr.res.in}

\date{\today}% It is always \today, today,
             %  but any date may be explicitly specified

\begin{abstract}
The kinetic analyses are quite important when it comes to understand the particle behavior in any device as they start to deviate from continuum nature. In the present study, kinetic simulations are performed using Particle-in-Cell (PIC) method to analyze the behavior of ions inside a cylindrical Inertial Electrostatic Confinement Fusion (IECF) device which is being developed as a tabletop neutron source. Here, the lighter ions, like deuterium are accelerated by applying an electrostatic field between the chamber wall (anode) and the cathode (cylindrical gridded wire), placed at the center of the device. The plasma potential profiles obtained from the simulated results indicate the formation of multiple potential well structures inside the cathode grid depending upon the applied cathode potential (from $-1$ to $-5~kV$). The ion density at the core region of the device is found to be of the order of $10^{16}~m^{-3}$, which closely resembles the experimental observations. Spatial variation of Ion Energy Distribution Function (IEDF) has been measured in order to observe the characteristics of ions at different cathode voltages. Finally, the simulated results are compared and found to be in good agreement with the experimental profiles. The present analysis can serve as a reference guide to optimize the technological parameters of the discharge process in IECF devices.
\end{abstract}

\maketitle

%\tableofcontents

\section{\label{sec:level1}Introduction}
Inertial Electrostatic Confinement Fusion (IECF) devices operate on the principle of the confinement of plasma particles in a purely electrostatic field. In such a device, ions are recirculated across the gridded cathode, confined and remain inside the system unless they collide with the neutrals, background ions and cathode grid wires. As a result, the secondary electron concentration also increases near the core region. The electrostatic potential produced by both the ions and electrons plays a vital role behind the fusion process in the system. The secondary electrons neutralize the space charge of ions and thus maintain the ion re-circulation across the gridded cathode \cite{murali2008study}. The highly energetic ions fused together at the core region and produce particles like neutron and proton. Unlike magnetic and laser-based confinement methods, which are primarily effective for long term power production, the IECF concept is being developed for the near term applications \cite{cipiti2003embedded,murali2008study}. The theoretical concept of IECF was first proposed by P. Farnsworth \cite{farnsworth1966electric}, and later, it was studied experimentally by R. Hirsch \cite{hirsch1967inertial} in 1970's. Different types of IECF devices have been developed till date, e.g., Hirsch \cite{hirsch1967inertial} introduced ion-gun injectors to confine ions, Nebel et al. \cite{nebel1995inertial} used a triple grid design for better confinement of the ions. Bussard \cite{bussard1994inherent} developed a magnetic-electrostatic version and other researchers came up with new versions such as single-grid \cite{miley1997discharge}, multiple-grid \cite{Dietrich2007,takamatsu2005inertial} magnetron assisted device etc. The potential profiles in the IECF device were first experimentally studied by Swanson et al. \cite{swanson1973potential} using electron beam probing. Later, Thorson et al. \cite{thorson1997convergence,thorson1998fusion} used emissive probe for the measurement of potential and ion density profile in the central region of the spherical IECF device by varying the gas pressure. Kipritidis et al. \cite{kipritidis2008absolute} have measured the fast hydrogen ion density in the vicinity of the cathode edge in a cylindrically symmetric IECF device using optical spectroscopy and Langmuir probe. Moreover, Yoshikawa et al. \cite{yoshikawa1999real} have carried out experiments for the direct measurement of potential by using laser-induced florescence technique. Over the years, researchers of different laboratories across the world (University of Wisconsin, University of Illinois, Tokyo Institute of Technology, Kyoto University etc.) \cite{murali2010effects,miley2005nuclear,murali2010consolidated,cipiti2005helium,boris2009novel,noborio2006confinement,yamauchi2001neutron,boris2009deuterium} have been continuously devoting their effort for upgrading the IECF device in terms of optimizing fusion reactivity, neutron production as well as its applications. As far as the theoretical studies are concerned, Nevin \cite{nevins1995can} has presented a model for the ion distribution function and is able to reproduce some of the essential features of the IECF system, like electrostatic confinement, strong ion peak at the center and a nearly mono-energetic distribution of the ions. Krall \cite{krall1992polywell} and Dolan \cite{dolan1994magnetic} introduced the polywell concept in the spherical IECF device in which potential well structures have been studied in order to achieve maximum fusion rate. The purpose of their work was to establish the IECF scheme as an alternate way to achieve thermonuclear fusion energy. However, later on it was confirmed that the IECF scheme shows little promise as a basis for the development of commercial electrical power plant \cite{nevins1995can}. Ohnishi et al. \cite{ohnishi1997correlation} observed the formation of potential well structure inside the cathode grid depending upon the magnitude of ion current. They have studied the dynamic behaviors  of potential well by performing numerical simulations on the basis of particle-in-cell (PIC) method. A correlation between the D-D neutron production rate and the depth of the potential well is also established in their work. Ohnishi et al. \cite{ohnishi2001particle} further improved the PIC code to enhance the computational speed and accuracy. However, they considered only the $D_{2}^{+}$ ions while $D^{+}$ ions are not included in the simulation. Again, the angular momentum that the ions acquire due to coulomb interaction and collision with the neutrals is not considered in the simulation. Buzarbaruah et al. \cite{buzarbaruah2015design} tried to track the trajectory of the $D^{+}$ ions by performing simulation using Simion code \cite{manura20168}. However, the limitation of the code did not allow to create the plasma environment inside the system, rather it could only provide the electric or magnetic field to observe the possible trajectories of ions.
\par 
The potential structure and ion density plays an important role for the occurrence of fusion reaction \cite{thorson1998fusion} (during higher voltage operations) in such devices, a detailed study of these profiles during lower cathode voltage conditions become equally important for a complete understanding of the underlying physics of the IECF scheme. In this work, we have studied the behavior of potential well and ion density profiles in the gas discharge plasma \cite{buzarbaruah2015design,buzarbaruah2017studyon,buzarbaruah2018study} during relatively lower cathode potential (up to $-5~kV$), in both $8$ and $16$ gridded system (see section 2 for details), using XOOPIC code \cite{verboncoeur1995object} (X11-based Object-Oriented Particle-In-Cell). The code is capable of producing the exact experimental scenario and all the essential parameters associated with the plasma can be extracted for analysis. Besides, the Ion Energy Distribution Function (IEDF) obtained at various locations of the simulated region for different cathode voltages is capable of providing us more information about the ions which is limited by the experiment. The simulated results then compared with the experiment to benchmark the results. It should be noted that here, we have basically focused our study to the structural and behavioral changes of the kinetic characteristics of the ions due to the application of potential in range $-1$ to $-5~kV$. However, fusion reaction of the particles does not occur in this applied potential range. In the next section, the experimental setup, procedures and the diagnostics used for the experiment are described. In section $3$, we briefly describe the simulation parameters and the code used in this simulation. The simulated results such as, potential well profile, ion density profile and the IEDF are described in the section $4$. The experimental results and their comparisons with the simulated results are also discussed in the same section. The last section of the paper contains the concluding remarks and the future scopes.

\section{\label{sec:level2}Experimental set up and procedure}
\begin{figure}
	\centering
	\includegraphics[width=8.0cm]{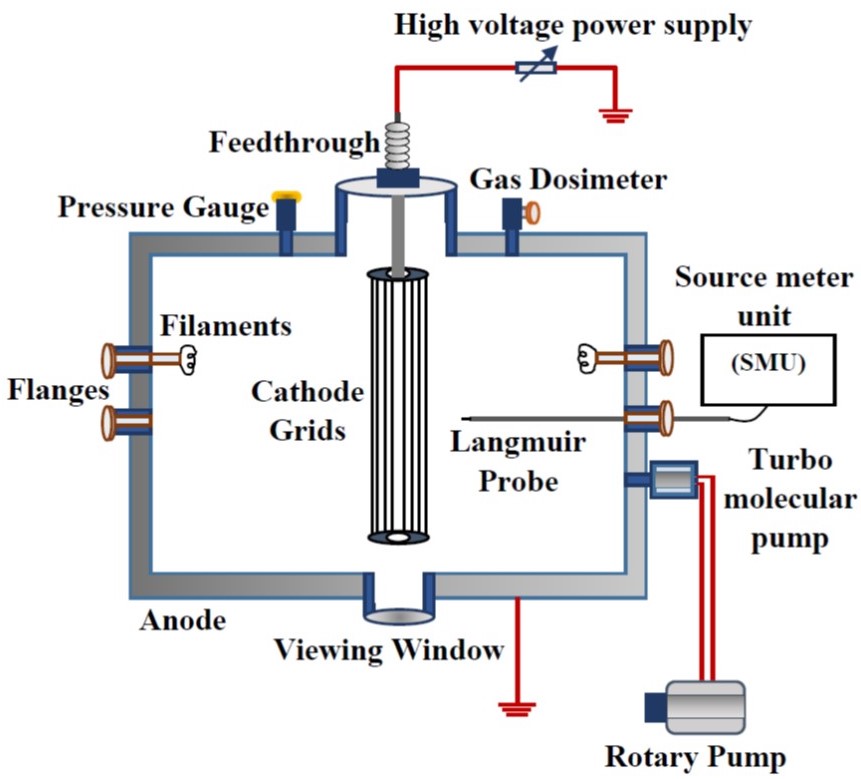}%
	\caption{Schematic diagram of the cylindrical IECF device with different accessories.
	\label{fig:schematic_diagram}}
\end{figure}

The IECF device used in this study consists of a cylindrical stainless steel vacuum chamber having a diameter of $50~cm$ and a height of $30~cm$. A highly transparent cylindrical cathode is kept vertically at the center of the chamber, which acts as the anode (grounded). The cathode that is connected to a high voltage power supply through a feedthrough, consists of tungsten grid wires of varying diameters. In this work, we have used two cathodes of diameter $3~cm$ each, made up of $8$ ($\sim92\%$ transparent) and $16$ ($\sim85\%$ transparent) numbers of grid wires, respectively. The dimension of the grid wires of both the cathodes are the same ($0.12~cm$ diameter). Different ports are there in the chamber for evacuation, viewing, coupling high voltage feedthrough, inserting gas and other diagnostic tools. A schematic diagram of the cylindrical IECF device used in the present work is shown in the figure(\ref{fig:schematic_diagram}). The chamber is evacuated by using a turbo molecular pump backed by a rotary pump and the pressure is maintained inside the chamber by a coarse feed valve.
\par 
The deuterium plasma is created by adopting hot cathode discharge (i.e. filamentary discharge) method in which two thoriated tungsten filaments are placed at two diagonally opposite positions and at $10~cm$ away from the wall of the chamber. The filaments are heated to produce thermionic electrons and a discharge voltage and current of $80~V$ and $200-500~mA$ is maintained, respectively, at a working pressure of $\sim10^{-3}$ Torr. Then, negative voltage is applied to the cathode through the feedthrough using a $5~kV$, $600~mA$ DC power supply. A cylindrical Langmuir probe of length $0.5~cm$ and diameter $0.05~cm$ has been used to characterize the plasma.
The probe is inserted into the plasma through one of the ports, as shown in the figure(\ref{fig:schematic_diagram}), and it is movable radially from the wall of the chamber to the center of the cathode. The potential at different positions inside the chamber is measured from the Langmuir probe. We have varied the applied cathode voltage from $-1$ to $-5~kV$ in order to observe the modifications in potential profile structure inside the cathode region \cite{bhattacharjee2019studies}. In order to measure the plasma temperature and hence the ion density, we have used a double Langmuir probe.

\section{Modeling}
\begin{figure}
	\centering
	\includegraphics[width=5.0cm]{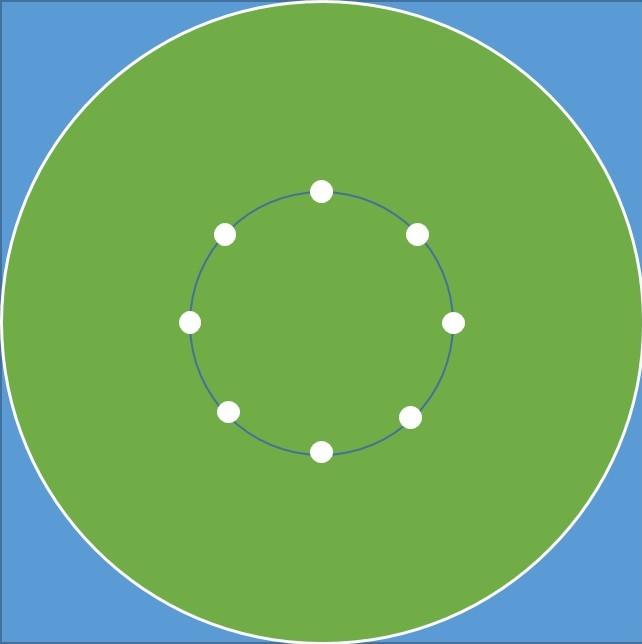}%
	\caption{Cross-section of simulation region with the cathode grid.
	\label{fig:cross_section}}
\end{figure}

To understand the complex behavior of ions inside the cylindrical IECF device, we have performed electrostatic particle-in-cell (PIC) simulation using XOOPIC code. It has the capability of handling two-dimensional space in cartesian and cylindrical geometries, including all three velocity components, with builtin Poisson solver. The code can deal with an arbitrary number of species and it includes Monte Carlo Collision (MCC) algorithms for modeling collisions of the charged particles with themselves and with the background gas. The simulated dimension and parameters are designed to recreate the actual experimental scenario in this paper. The simulation geometry is a horizontal $2D$ cross-section of the device including the cross-section of the cathode grids, (figure(\ref{fig:cross_section})) as per the experimental setup. We have modeled electron emitters inside the simulation domain, each emitting a constant and equal flux of electrons continuously into the system. The emitted electrons interact with the background gas (deuterium) and produce ions due to ionizing collision. A high negative voltage is applied to the cathode grid wires, which accelerates the ions towards it. The system is assumed to be bounded and symmetric. The time step for the simulation has been chosen satisfying the Courant condition \cite{de2013courant} and ensuring particles must not fly over more than one cell in one time step. The size of the cell or the spatial step $\Delta x$ is considered in such a way that it is always smaller or equal to the Debye length of the system for different operating conditions. To ensure no particles are moving more than one cell length per time step,  the time step size is resolved based on the fastest particle present in the system. The drift velocity ($v_d$) of the fastest particle will decide the time-step size ($\Delta t$) and it is given by \cite{birdsall2004plasma}.
         \begin{equation}
          \Delta t=0.3\times\frac{d}{v_d}~;~~d = \frac{1}{\sqrt{(\frac{1}{dx})^{2}+(\frac{1}{dy})^{2}}} \label{eq1}
         \end{equation} 
where, $dx$ and $dy$ are the cell sizes in the $x$ and $y$ directions, respectively. Putting appropriate values in the above equation, the $\Delta t$ is found to be of the order of $10^{-11}~s$. As far as collisions are concerned, XOOPIC includes a Monte Carlo collision model which can handle non-interacting gas mixtures, including elastic, excitation, ionization and charge exchange collisions \cite{becker2004non}. However, keeping the background gas pressure constant, any recombination process is ruled out. The absorbing boundaries balance out any loss of charge particles due to recombination \cite{becker2004non}. Multigrid Poisson solver is used to solve the Poisson's equation which also takes care of the boundary conditions of the simulation. The reason behind the choice of multigrid solver over others is it's basic principle of solving Poisson's equation first on a coarse grid, and then using the solution as a guess, to solve the equation again for a finer grid. It speeds up the solution for the finer grid, reducing the total computational time required to converge to an accurate solution. Dirichlet boundary conditions are imposed by the solver on the Cartesian grids of the regular domain boundaries. The input file of the XOOPIC simulation is developed by taking into account all the parameters and conditions mentioned above. In order to get a good resolution of phase space, it requires an extremely large number of computational particles (super particles). However, considering so, it is important to optimize the number as it significantly influences the total run-time.  

\begin{table}
	\caption{Simulation parameters} 
	\label{Table 1}
	\begin{center}
	\footnotesize
    \begin{tabular}{@{}ll}
		\textbf{Parameters} & \textbf{Values} \\
		Grid size & $512\times512$ \\ 
		%	Number of cells & $10,816$ \\ 
		Length & $0.21~m$ \\  
		Width & $0.21~m$ \\ 
		Time step (dt) & $10^{-11}~s$ \\
		Specific weight & $10^{9}$ \\
		%	Number of particles ($N_{p}$) & $10^{5}$ \\
		Background gas & Deuterium \\
		Cathode potential & $-1~kV$ to $-5~kV$ \\
		Anode potential & $0~V$ \\
		\end{tabular} 
	\end{center}
	 \end{table}%
\normalsize
In PIC simulation, the trajectory of every super-particle is computed kinetically. Therefore, longer domain length with a large number of super particles requires much longer run time to solve the problem. Again, keeping the total number of macro-particle same, if the simulation domain length is increased, the number of simulated macro-particle per Debye sphere is decreased at the same rate, which will eventually increase the statistical noise associated with the simulation. To avoid this, a large number of computational particles with relatively shorter domain length is appropriate for the simulation to obtain the desired results. The code is capable of running in both GUI (Graphical User Interface) and non-GUI modes. To reduce the execution time, the non-GUI mode is generally preferred.
\begin{figure}
	\centering
	\includegraphics[width=6.5cm]{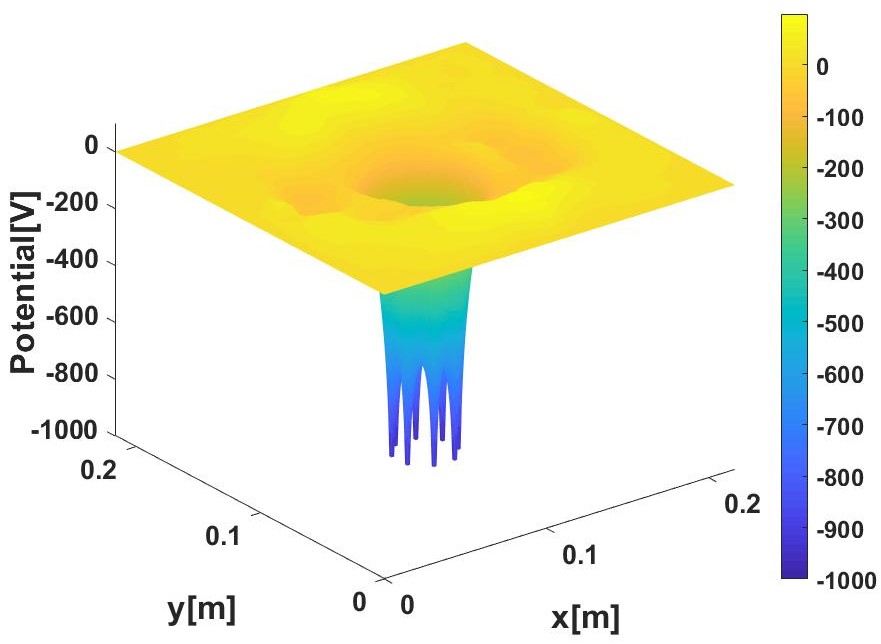}
	\caption{Surface plot of potential profile at $-1~kV$ cathode voltage.
	\label{fig:surf_plot}}
\end{figure}

\begin{figure}
% 	\begingroup
	\centering
	\subfloat[]{
	\includegraphics[width=6.5cm]{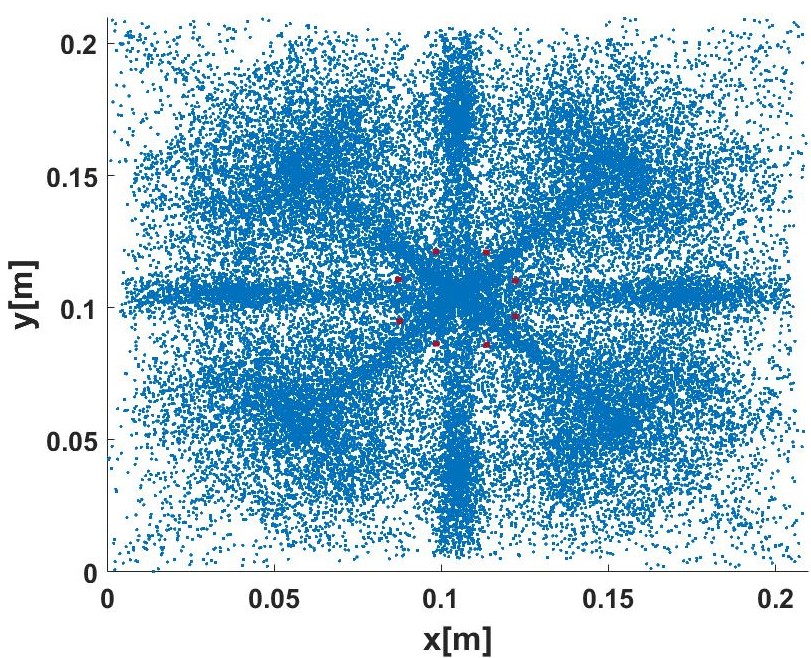}
	}\hfill
	\subfloat[]{
	\includegraphics[width=6.5cm]{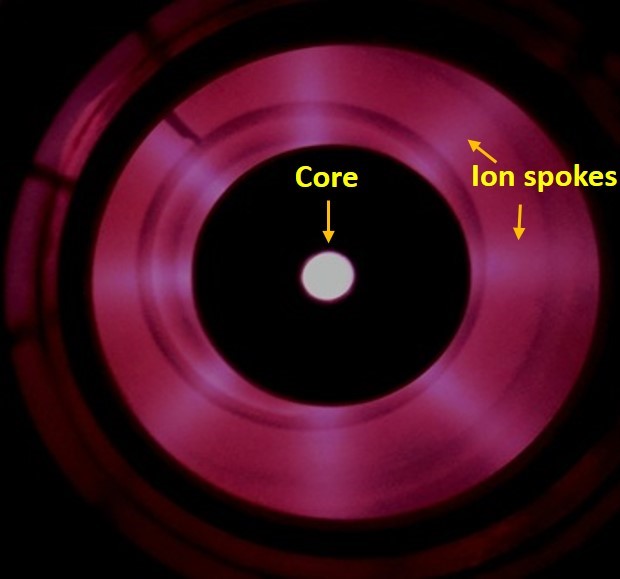}
	}
% 	\endgroup
	\caption{Observation of re-circulation of ions in both simulated profile (a) and in (b) experiment.
	\label{fig:recirc_both}}
\end{figure}

% \begin{figure}
%     \centering
% 	\begin{subfigure}{7cm}
% 	\centering
%     {\includegraphics[width=\linewidth]{4(a).jpg}}
%     \caption{}
%     \end{subfigure}
%     \begin{subfigure}{6.5cm}
% 	\centering
%     {\includegraphics[width=\linewidth]{4(b).jpg}}
%     \caption{}
%     \end{subfigure}
% 	\caption{Observation of re-circulation of ions in both simulated profile (a) and in (b) experiment.
% 	\label{fig:recirc_both}}
% \end{figure}
For such simulations with source involved, saturation in the total number of macro-particles in the system indicates the steady-state. A typical computer run takes at least $2-3$ days to achieve the steady-state. After reaching the steady-state, the diagnostics are saved as ASCII data files. The simulation parameters used in this work are depicted in the table \ref{Table 1}. The primary electron and ion temperatures are assumed to be $3$ and $0.1~eV$, respectively. Along with the simulations, a few scripts have been developed in Python and MATLAB\textsuperscript{\textregistered} to visualize the data. 
\begin{video}
\href{https://zenodo.org/record/4290965/files/animation.mp4}{\includegraphics[width=6.5cm]{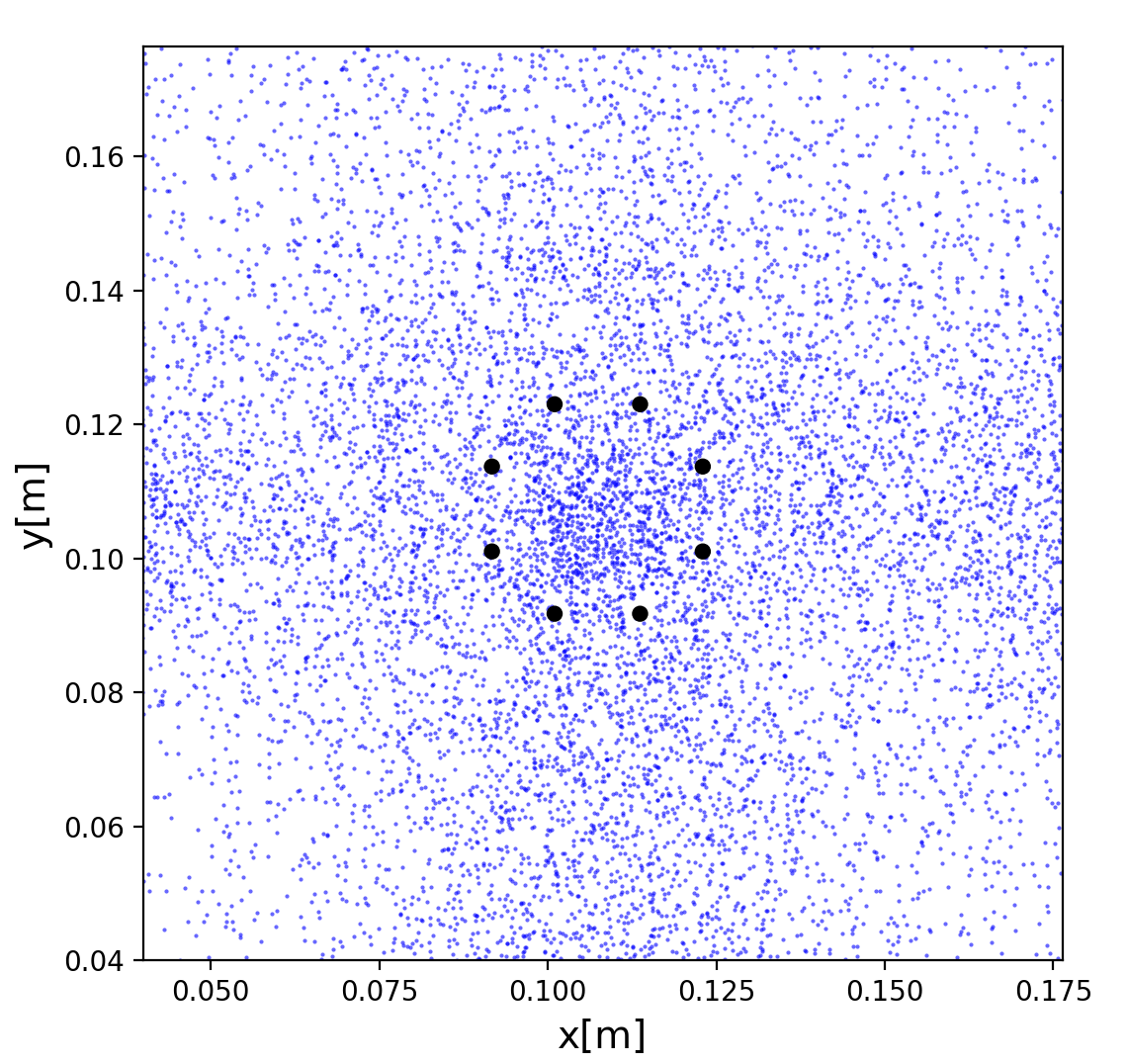}}%
 \setfloatlink{https://doi.org/10.5281/zenodo.4290965}%
 \caption{\label{vid:phaserecord}%
 The snapshot from the ion phase space structure recorded during $-1~kV$ cathode voltage indicating the re-circulation of ions across the cathode grid openings.
 }%
\end{video}
\section{Results and discussions}
The re-circulation of the ions across the cathode grid openings is the fundamental process of the IECF scheme. Here, the energetic ions collide with themselves and with the background gas and take part in fusion process at high applied voltages (above $-30~kV$). The re-circulation of the ions can be visualized from the movement of the ions in the phase space during run-time (see video \ref{vid:phaserecord} or the supplemental material \cite{supplimat}). One such phase space plot of ions during $-1~kV$ simulation is shown in the figure(\ref{fig:recirc_both}{a}). The term re-circulation signifies the to and fro motion of the ions across the central grid due to the applied electrostatic potential. As the negative potential is applied on the cathode the ion will immediately rush towards it and move towards the other side of the chamber through the cathode grid openings due to its inertia. However, the electrostatic force due to the negative grid biasing will slow it down and after reaching the extreme position the ion will eventually come back towards the cathode grid and continues the same circular motion in different paths unless it collides with the grid wires. In due process, some of the ions get trapped inside the cathode, and their density is found to be increasing at the central core region. The increase in ion density inside the cathode mainly depends upon the applied cathode voltage and the current. Figure(\ref{fig:recirc_both}{b}) shows the photograph taken from the bottom side of the cylindrical IECF device during $-1~kV$ cold cathode discharge experiment, in which spokes of ions are coming out of the core region of the cathode. 
\begin{figure*}
	\begingroup
	\centering	
		\subfloat[]{\includegraphics[width=6.5cm]{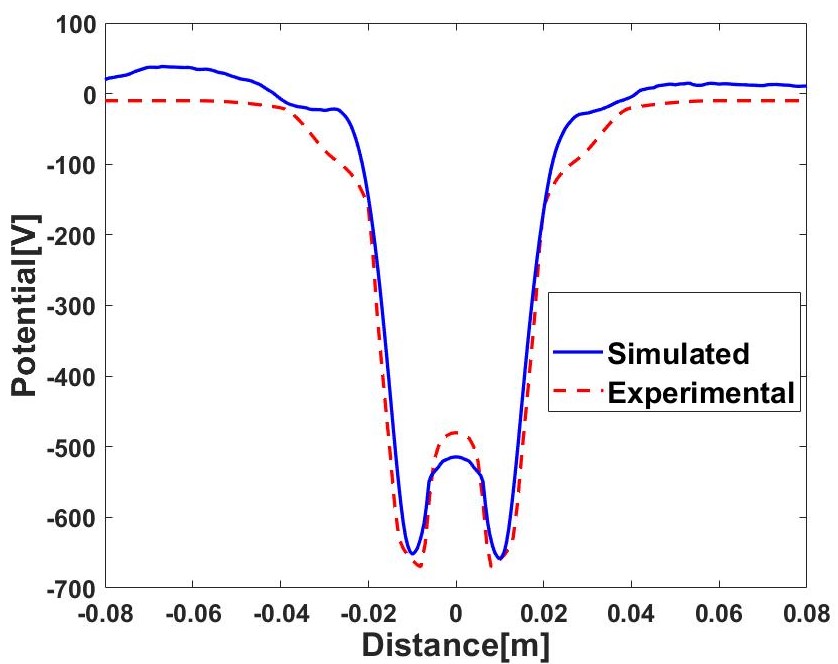}}
		\subfloat[]{\includegraphics[width=6.6cm]{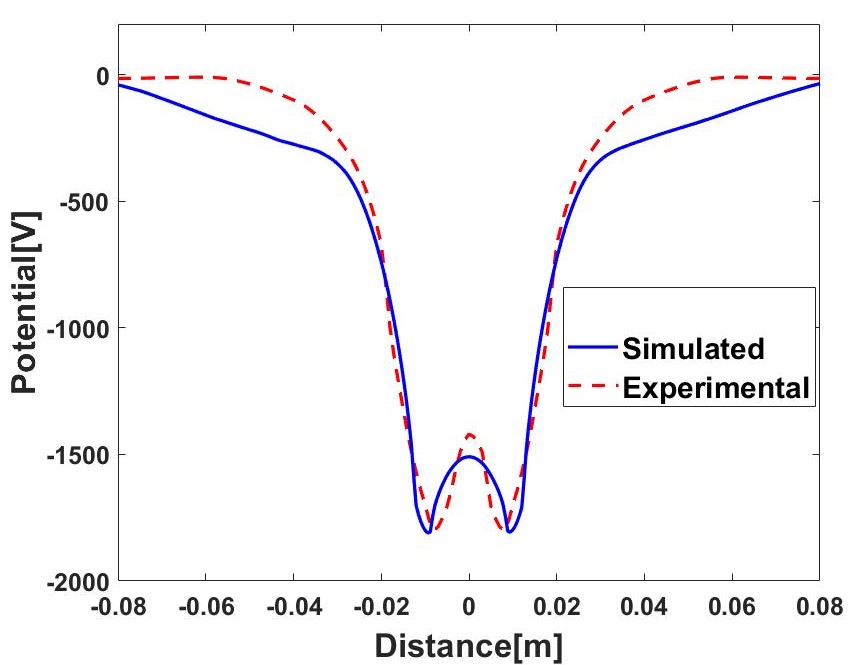}}\\
		\subfloat[]{\includegraphics[width=6.5cm]{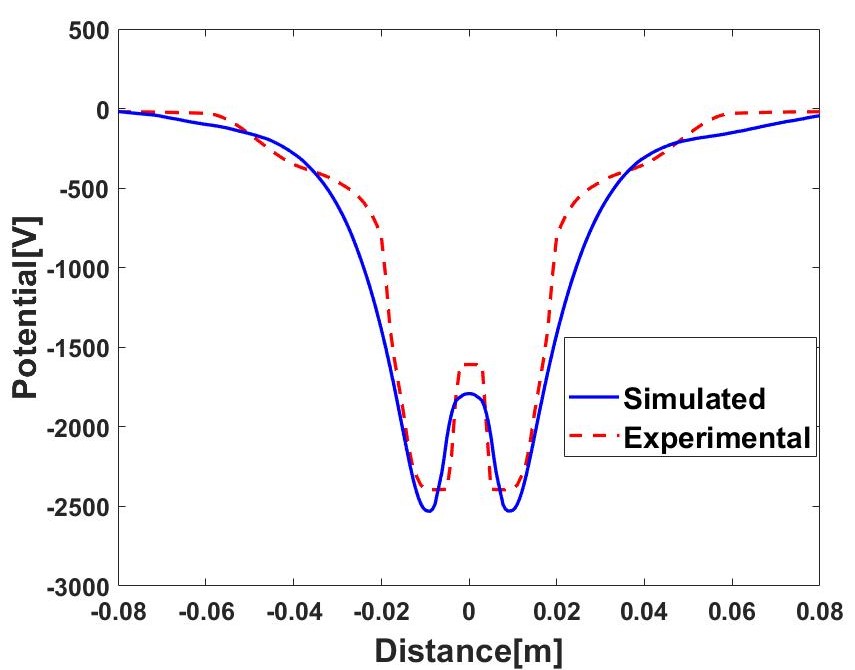}}
		\subfloat[]{\includegraphics[width=6.5cm]{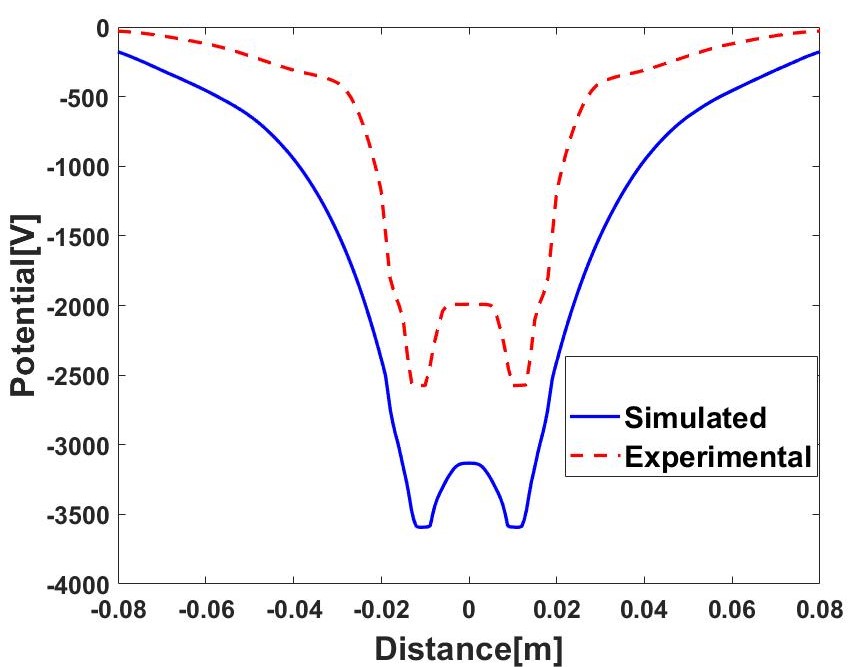}}\\
		\subfloat[]{\includegraphics[width=6.5cm]{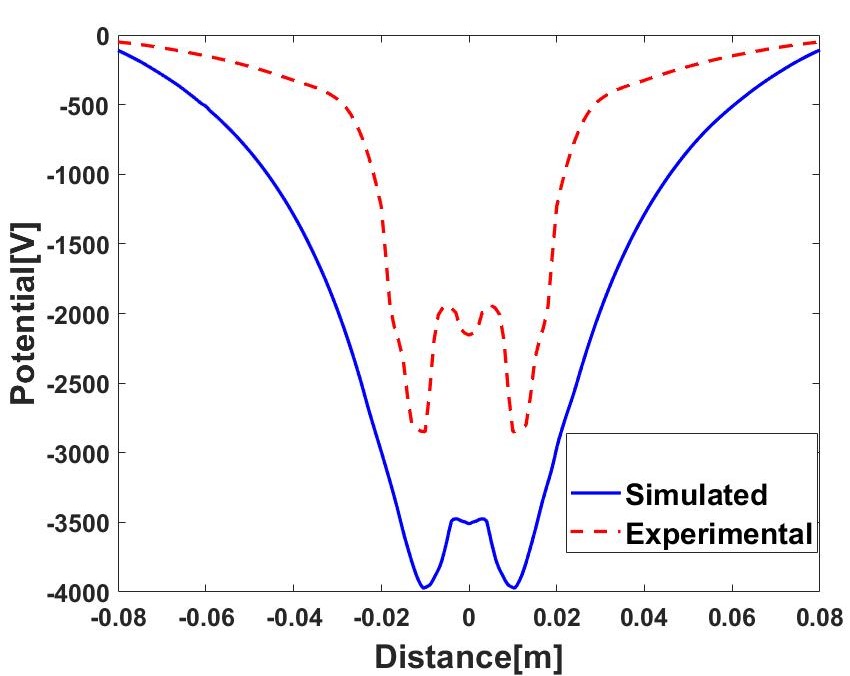}}
	\endgroup
	\caption{Simulated potential profiles (blue lines) compared with the experimental ones (dashed red lines) during (a) $-1~kV$, (b) $-2~kV$, (c) $-3~kV$, (d) $-4~kV$ and (e) $-5~kV$ cathode voltage, respectively.\label{pot_1}}
\end{figure*}
Both the simulated plot and the experimental photograph displays the dominating channels of ion motion across the cathode openings in the IECF system. This mode of operation is popularly known as the star mode \cite{miley2014inertial} in IECF system. The surface plot of potential obtained from the simulation data during $-1~kV$ applied cathode voltage is shown in figure(\ref{fig:surf_plot}). A detailed discussion of the potential and ion density profiles are done in the next subsection.

\subsection{Potential and ion density profiles}
Voltage ranging from $-1$ to $-5~kV$ is applied to the gridded cathode (having $8$ numbers of grid wires) of diameter $3~cm$ to observe the modifications in the potential profiles, specially inside the cathode region. After reaching the steady-state in the simulation, the obtained potential profiles are plotted for better analysis. The equatorial plots of the potential profiles for different cathode voltages are compared with the experimentally established results \cite{bhattacharjee2019studies}, as shown in figure(\ref{pot_1}). For lower cathode voltages, the ions re-circulate across the cathode openings until they collide with themselves or with the cathode grid. This results in scattering of the ions out of the potential trap. As the cathode voltage is increased, e.g. at $-1~kV$, the ion flux significantly increases inside the cathode, which results in the formation of a space charge of ions, i.e., a virtual anode inside the cathode (figure(\ref{pot_1}{a})). Similar results are obtained during $-2~kV$ and $-3~kV$ cathode voltage operations, as shown in figures(\ref{pot_1}{b}) and (\ref{pot_1}{c}), respectively. The depth of the potential well is also found to be increasing with the increase in cathode voltage. The formation of virtual anode reflects back further incoming ions even before reach the center.
			
However, the virtual anode serves as the potential trap for the secondary electrons (emitted from the gridded cathode due to the collision of ions) and they oscillate inside the virtual anode just like the ions in the outer trap. If the electron density increases, they in turn form a space charge of electrons which may lead to the formation of another virtual electrode, i.e., a virtual cathode inside the virtual anode. In principle, the process would continue to form multiple numbers of virtual electrodes inside the real cathode, but in practice, the potential trap formed by the secondary electrons, i.e., the first virtual cathode is all that is observed till date \cite{murali2008study,miley2014inertial}. The experimental results are in good agreement with the simulated results up to $-3~kV$ cathode voltage, as shown in figure(\ref{pot_1}). The depth of the potential well increases as we increase the cathode potential and a prominent virtual anode at the center of the cathode grid have also been observed in the experimental plots. The structures of the potential profiles for $-4~kV$ and $-5~kV$, in experimental cases are seen to differ from the simulated ones. This variation is due to the Debye shielding of the Langmuir probe (experimental case) when operated in such high potential regions. Moreover, insertion of a probe inside the chamber creates a disturbance in the ion sheath region and therefore a drop in the sheath potential is observed \cite{thorson1997convergence}. Therefore, exact measurement of the potential values is much difficult to obtain. On the other hand, for simulations no such physical probe is required to measure plasma potential hence, disturbances are not accounted. In $-4~kV$ simulation, potential drop of $10\%$ ($\sim-3600~V$) is observed while it is about $35\%$ ($\sim-2600~V$) in the experiment. Similarly, for $-5~kV$ applied potential the drop is about $20\%$ in the simulated profile and $44\%$ in the experimental one. During $-5~kV$ cathode voltage, a clear formation of a virtual cathode has been observed inside the cathode grid, experimentally \cite{bhattacharjee2019studies}, while in the simulated result (figure(\ref{pot_1}{d})), only a slight indication of the virtual cathode is observed at the center. This is due to the numerical limitations specially during the simulation of higher voltage operations (around $-5~kV$). The measurement of small potential variations inside the cathode is one of the difficult tasks since the grid (computational) resolution is inadequate. Such problem might be overcome with better grid resolution and higher number of macro-particles. The potential profiles presented here in this paper are the time averaged data of the specific configurations. For high voltage operation e.g.  $-5~kV$, due to the additional trapping of electrons inside virtual anode, determining the accurate profile becomes more complicated.  
\begin{figure}
    \centering
    \includegraphics[width=6.5cm]{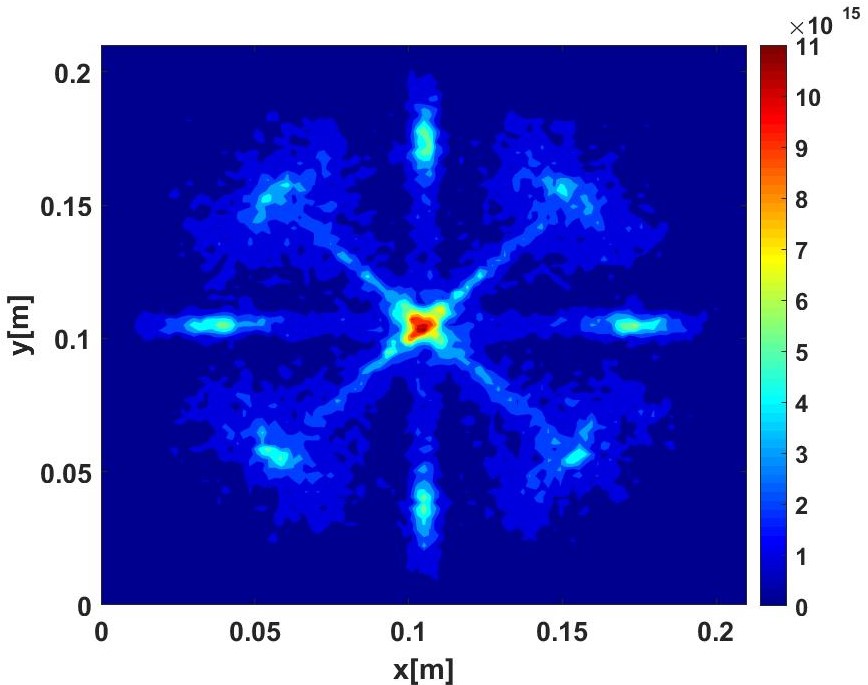}
    \caption{Contour plot of ion density at $-1~kV$ cathode voltage.}
    \label{fig:contour}
\end{figure}

Therefore, in such cases, the signature of virtual electrode formation as seen in figure(\ref{pot_1}{d}) is believed to be the proper justification of the detected profiles in the experiment.       
\begin{figure*}
    \begingroup
		\centering
		\subfloat[]{\includegraphics[width=6.5cm]{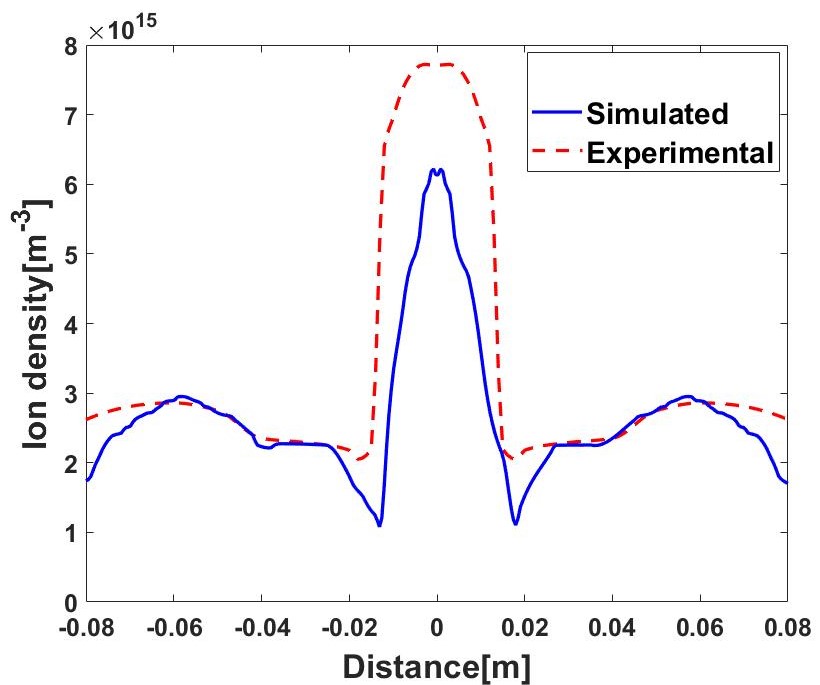}}
		\subfloat[]{\includegraphics[width=6.5cm]{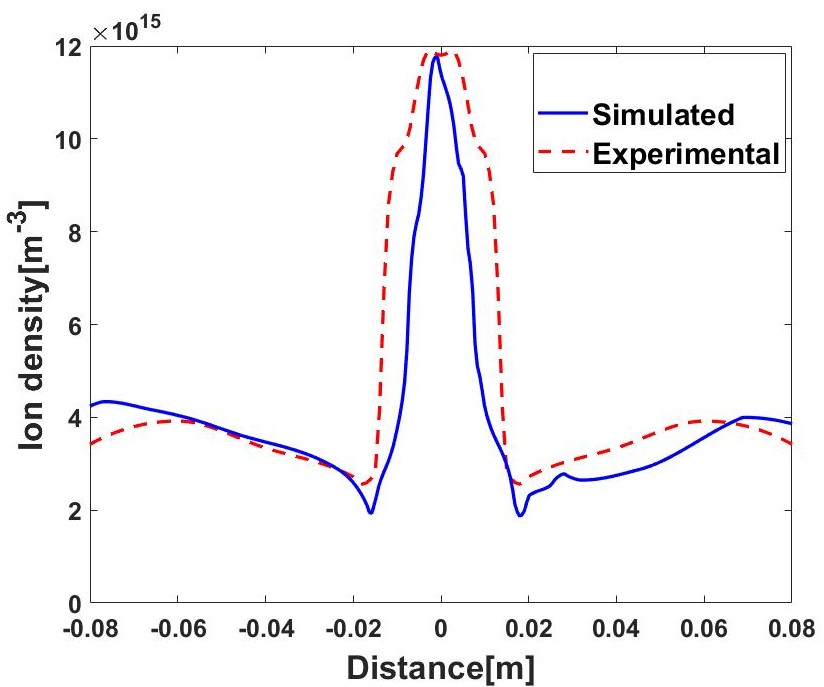}}\\
		\subfloat[]{\includegraphics[width=6.5cm]{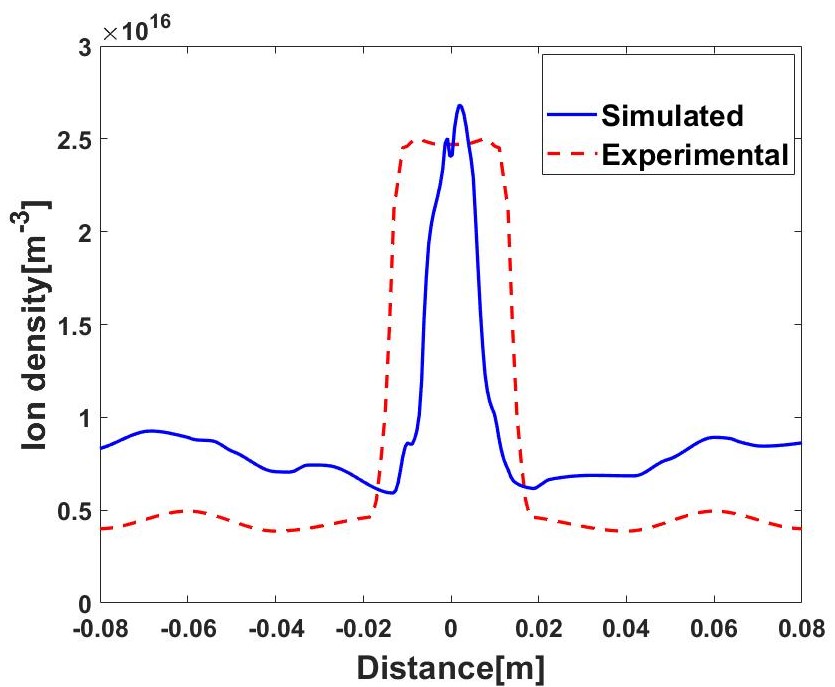}}
		\subfloat[]{\includegraphics[width=6.5cm]{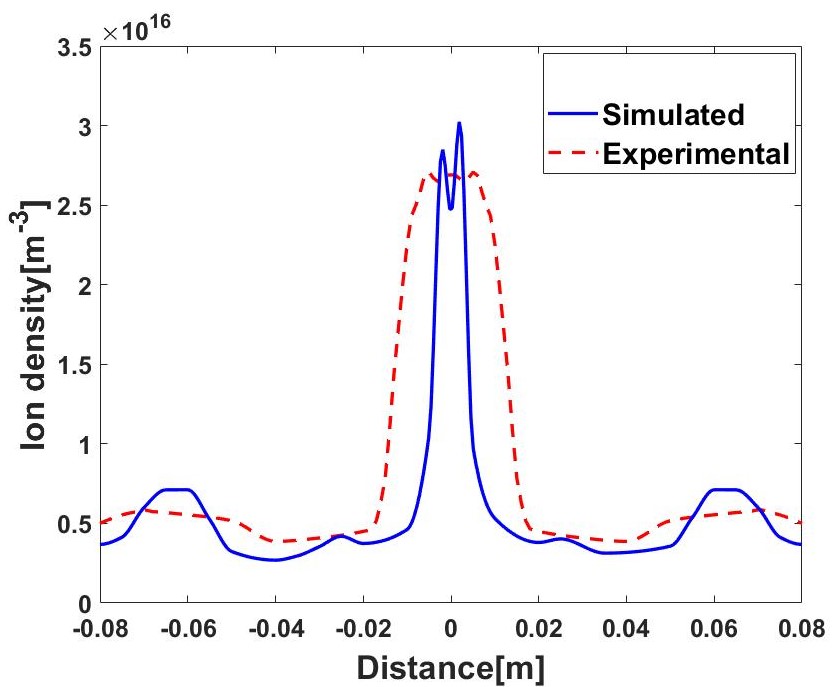}}
	\endgroup
	\caption{Simulated ion density profiles (blue lines) compared with the experimental ones (dashed red lines) during (a) $-1~kV$, (b) $-2~kV$, (c)$-3~kV$ and (d) $-4~kV$ cathode voltage, respectively.\label{ion_profile}}
\end{figure*}
\par As already mentioned, ion density also plays a crucial role for carrying out the fusion process and to produce neutrons from such devices. The contour plot of the ion density at $-1~kV$ cathode voltage operation is depicted in figure(\ref{fig:contour}). We have studied the simulated ion density profiles up to a cathode voltage of $-4~kV$ and are shown in figure(\ref{ion_profile}). An abrupt increase in ion density has been observed inside the cathode because of the trapped ions. At $-1~kV$ cathode voltage, the maximum density observed is $6.3\times10^{15} m^{-3}$, as shown in figure(\ref{ion_profile}{a}). The density tends to be increasing as we gradually increase the cathode voltage, and at $-4~kV$ (figure(\ref{ion_profile}{d})) it is found to be one order higher ($3.1\times10^{16} m^{-3}$) than the first case. In all the cases, ion density outside the cathode, ($0.04$ to $0.06~m$ from the center) in both sides from the center, is observed to be slightly increasing. This is due to the re-circulating nature of the ions from one side to the other (figure(\ref{fig:recirc_both}{a})). During the re-circulation process, the ions reaching the extreme position at one end achieve minimum velocity and turn back towards the negatively biased cathode with increased velocity. At the position of minimum velocity the density of ions is found to be increasing, which satisfies the ion continuity. A slight increase in ion density outside the cathode region is also observed in the experimental results (figure(\ref{ion_profile})). Both the simulated and the experimental profiles indicate that the ion density at the core region increases with the applied cathode voltage and it is expected to be increasing further with higher cathode voltage. However, the experimental density profiles are found to be wider than the simulated profiles inside the cathode region. The discrepancy can be explained as follows. In experiments, we have measured the ion density at a spatial resolution of $2~mm$ by moving the Langmuir probe manually from chamber wall to the core plasma region. The tip of the probe is vertically attached to the probe holder (SS rod). While moving the probe inside the plasma region, the probe accesses the plasma throughout its tip length and accordingly records the data. Therefore, due to the active measurement by a physical probe having $5~mm$ length, such discrepancies appear while comparing with the simulated results.
 \begin{figure*}
    \begingroup
		\centering
		\subfloat[]{\includegraphics[width=6.5cm]{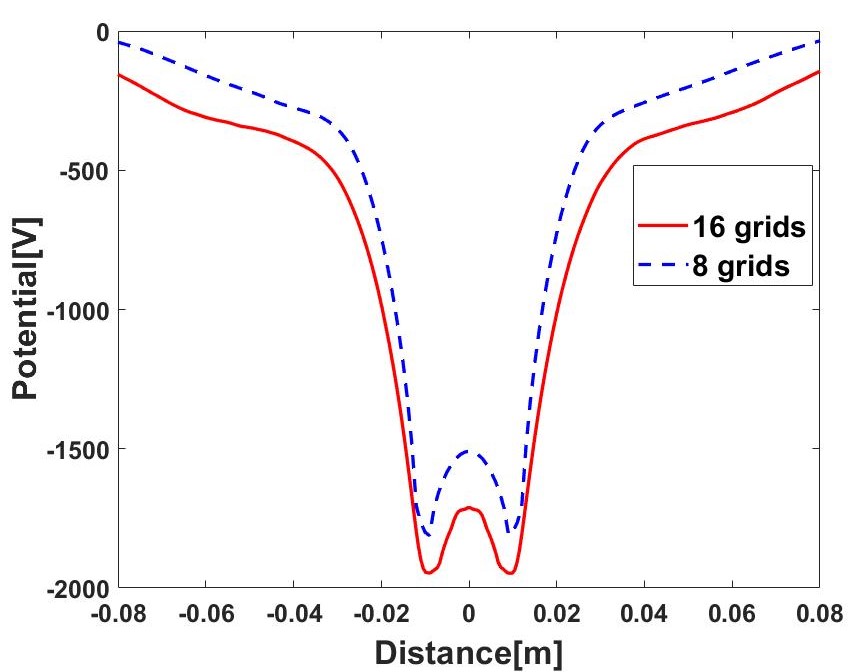}}
		\subfloat[]{\includegraphics[width=6.5cm]{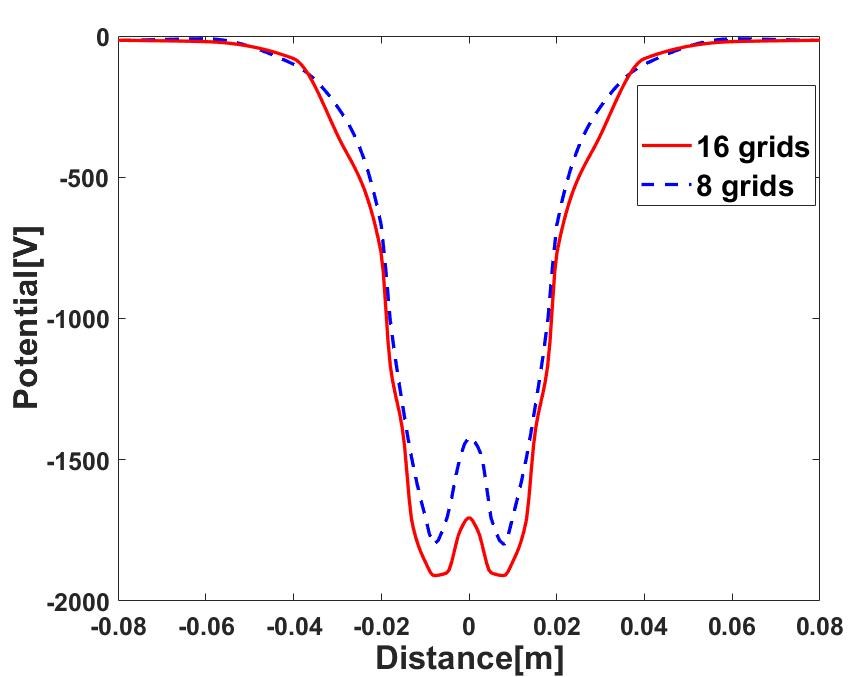}}
	\endgroup
	\caption{Comparison of potential profile of $8$ (dashed blue lines) and $16$ gridded (red lines) cathode in both (a) simulated and (b) experimental plots during $-2~kV$ cathode voltage operation.\label{pot_16grds}}
\end{figure*}
\begin{figure*}
    \begingroup
	\centering	
	\subfloat[]{\includegraphics[width=6.5cm]{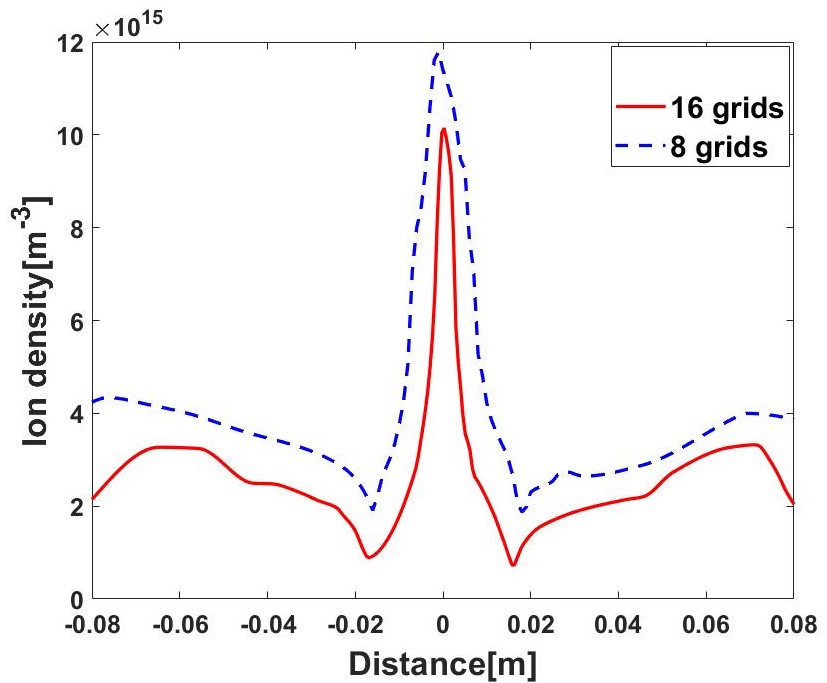}}
	\subfloat[]{\includegraphics[width=6.5cm]{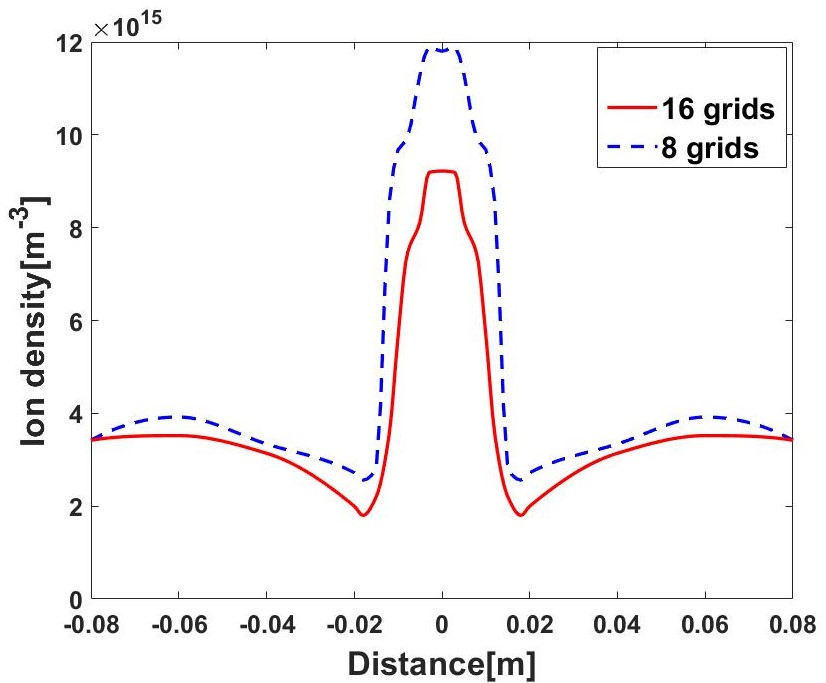}}
	\endgroup
	\caption{Comparison of ion density profile of $8$ (dashed blue lines) and $16$ gridded (red lines) cathode in both (a) simulated and (b) experimental plots during $-2~kV$ cathode voltage operation.\label{den_16grds}}
\end{figure*}

\par {We have performed another set of simulations to study the effect of grid transparency on potential well structure and ion density profiles by changing the number of cathode grid wires to $16$, keeping the cathode diameter same. We have kept all the conditions same as described in $8$ gridded cathode and the results are shown in figure(\ref{pot_16grds}), and (\ref{den_16grds}). This time we have shown the results of only $-2~kV$ operation rather than displaying all the results. Here, in the simulated potential profile, the maximum negative potential is observed to be higher than the $8$ gridded system. It can be explained as: the measurements have been done along the line which passes through the mid point of the two consecutive cathode grid wires situated on either sides of this line. The mid point between the two nearby grid wires on either sides of the line of measurement gives us the maximum negative potential values. It can be easily visualize that the spacing between the cathode grid wires is lesser in a $16$ gridded system than that in a $8$ gridded system having the same diameter. Therefore, the line of measurement is more nearer to the consecutive grid wires in case of a $16$ gridded system, giving the maximum negative potential value slightly higher (approaching the applied cathode voltage, $-2~kV$) as compared to the $8$ gridded system, as depicted in figure(\ref{pot_16grds}{a}). Experimental plot (figure(\ref{pot_16grds}{b})) also shows similar behavior. Due to similar reason the ion density is found to be minimum near the grid wires in $16$ gridded system as compared to the $8$ gridded one.
\begin{figure*}
    \begingroup
	\centering
	\subfloat[]{\includegraphics[width=6.5cm]{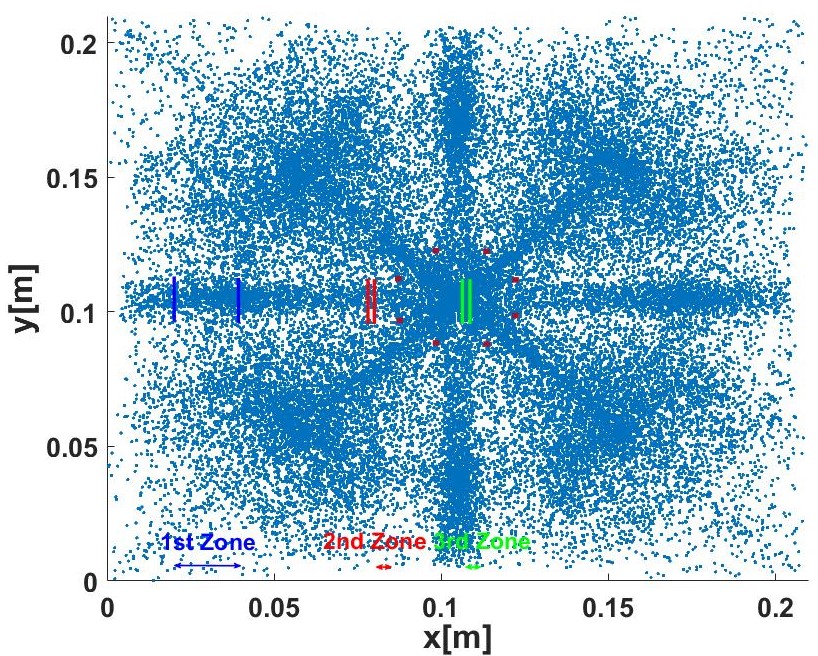}}
	\subfloat[]{\includegraphics[width=6.5cm]{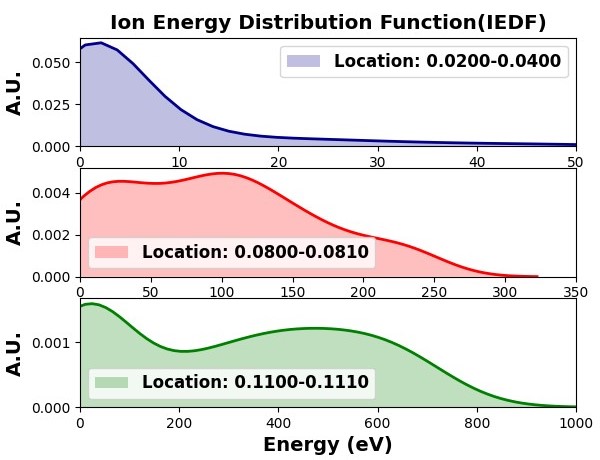}}\\
	\subfloat[]{\includegraphics[width=6.5cm]{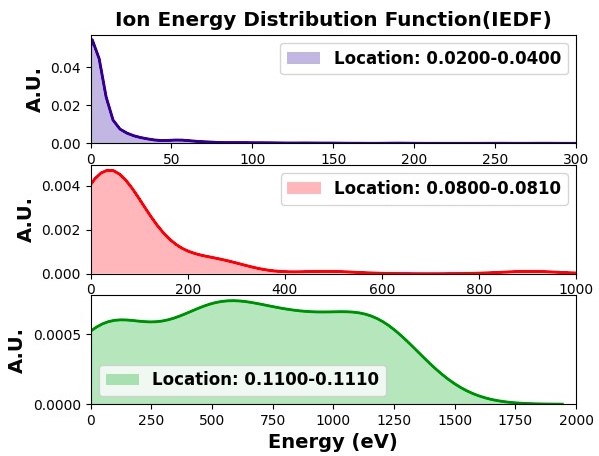}}
	\subfloat[]{\includegraphics[width=6.5cm]{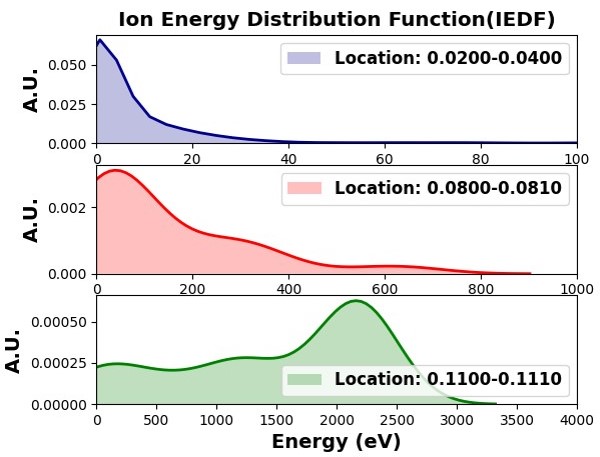}}
	\endgroup
	\caption{Different positions (zones) at which IEDF is measured is shown in figure (a). Simulated profiles of IEDF at cathode voltages $-1$, $-2$, and $-3~kV$ along x-component of velocity in three different zones are shown in figures(b), (c), and (d), respectively. \label{fig:iedf}}
\end{figure*}
Another important case is also noticed that the peak ion density inside the $16$ gridded cathode is lesser than that of the $8$ gridded one as observed in both the simulated (figure(\ref{den_16grds}{a})) and experimental (figure(\ref{den_16grds}{b})) profiles. Geometrical transparency of the cathode plays a vital role in this case. The ion density at the core region is directly related to the transparency (spacing between the grid wires) of the cathode \cite{thorson1997convergence,miley2014inertial}. Therefore, $16$ gridded cathode which is geometrically less transparent ($\sim85\%$) than the $8$ gridded one ($\sim92\%$), we have observed relatively lower ion density at the core.}
\subsection{Ion energy distribution function (IEDF)}
In order to measure the ion energy distribution from the simulation data, we have prepared a Python script which provides us the IEDF profiles at different locations of the simulation domain \cite{schwager1990collector,adhikari2018ion}. We have chosen three different locations, one inside and two outside the $8$ gridded cathode, as shown in figure(\ref{fig:iedf}{a}), where IEDF's are measured during different cathode voltages. IEDF is measured during three different cathode voltages, $-1$, $-2$ and $-3~kV$, as shown in figure(\ref{fig:iedf}). Since the device is symmetrical along both x and y directions, the ions along y-component of velocity show similar type of characteristics as that of the ions along x-component. The energies are measured in eV and the y-axis is provided in Arbitrary Units (A.U.) from Gaussian Kernel Density Estimate.
In case of $-1~kV$ cathode voltage, single peak (blue curve) in the IEDF profile outside the cathode ($1^{st}$ zone) suggests these regions are populated with low energy ions (primary ions due to ionization \cite{lieberman2005principles,liu2012main}) only. A second peak, in addition to the primary peak, is observed in the $2^{nd}$ zone of the IEDF (red curve). The energy of the ions re-circulating across the cathode grids can be described by the second peak. Since, the re-circulation of ions is dependent on the applied potential, the energy in this zone can vary based on the potential. If we compare the red curves for all three cases, such mechanism can be understood. Inside the cathode region ($3^{rd}$ zone), which is the area of interest, shows two distinct peaks (green curve).  One at the tail of the distribution along with the primary peak, as shown in figure(\ref{fig:iedf}{b}). The ions which are trapped inside the potential well will have another particular range of energy (or frequency) than the ions outside the trap. These ions can be represented by the peak in the third zone at the high energy tail of the distribution function. Similar peaks are also observed at different energy ranges in case of $-2$ and $-3~kV$ cathode voltage operations, as shown in figures(\ref{fig:iedf}{c}) and (\ref{fig:iedf}{d}), respectively. Figure \ref{fig:iedf} gives us an overview of the kinetic characteristics ions inside IECF device and its variability due to applied potential. From the IEDF profiles in the third zone for all the cases, we can observe the shift from a double Gaussian distribution \cite{kurt2016particle} to a single Gaussian distribution populated with trapped ions. Higher applied potential allows more ions to be trapped in the potential well, hence the percentage of recirculated ions get reduced.
\section{Conclusion}
The results obtained through simulation, in this paper, suggest that the potential profiles are in good agreement with the earlier obtained experimental profiles during applied cathode potential ranging from $-1$ to $-5~kV$. The depth of the potential well increases with the cathode voltage. An indication of the formation of multiple potential wells is also observed in the simulated profile during $-5~kV$ cathode voltage in the $8$ gridded system, while the experimental profile shows more prominent structure. Similarly, simulated profiles of ion density are also matching well with the experimental profiles. Lastly, the IEDF inside the cathode shows a transition from double Gaussian to single Gaussian, which suggest that the ions at the center oscillate with different frequencies end up with a different sets of energies inside the potential well during high voltage operations and the population of such ions increases with applied potential. The low energy peak for smaller applied potential represent the ions which are re-circulating across the openings of the cathode grids and the high energy peak corresponds to the ions, getting trapped at the core region. As far as the future scopes are concerned, further improvement of the domain configuration is needed in order to perform the simulation for higher voltage operations and to establish a concrete evidence of the formation of multiple potential wells inside the cathode. Due to its serial nature, the present code can run on a single CPU with a limit on the maximum allowed computational grid resolution, which we found to be insufficient to resolve the cases for higher applied cathode voltages. The possible solution to overcome the issue would be either to adopt finite element serial PIC codes or massively parallel finite-difference PIC codes. The finite element PIC codes would give us more flexibility to increase the number of cells around the cathode grids and subsequently reducing the number of cells outside the grid. It would allow us to perform high-resolution simulations without extra computational load. In comparison, the massively parallel finite-difference PIC codes would allow us to use a larger simulation domain with an increased number of macro-particles to resolve the presently facing issue. We are currently exploring both the options and confident of reporting a more complex study with high voltage operations in the near future. 
\par The discharge process and the success of IECF device as a neutron source highly depend on the understanding of  potential structure as well as on the ion dynamics inside the device. The present analyses are believed to serve as a reference guide to optimize technological parameters in IECF devices.

\begin{acknowledgments}
The authors are grateful to the Director, Institute for Plasma Research (IPR), Gandhinagar, India and the Center Director, Center of Plasma Physics-Institute for Plasma Research (CPP-IPR), Sonapur, India, for providing us the opportunity to carry out this work. We are thankful to Prof. H. Bailung, IASST, Guwahati, India, for his valuable suggestions. We also thank Mr. M.K.D. Sarma for his technical support.
\end{acknowledgments}

% \bibliography{manuscript}% Produces the bibliography via BibTeX.

%%%%%%%%%%%%%%%%%%%%%%%%%%%%%%%%%%%

%

\end{document}